\documentclass[a4paper, 11pt]{article}
\usepackage{jinstpub} 

\usepackage[version=4,
arrows=pgf-filled,]{mhchem} 

\usepackage{textgreek}
\usepackage{siunitx} 
\usepackage{trimclip}
\usepackage[noabbrev]{cleveref} 

\title{Laboratory and Synchrotron Validation of \textmu-XRF for Sulfur Mapping in CTMP Paper Samples}

\author[a,b,1]{F. Foroughi,\note{Corresponding author.}}
\author[a]{H. Bergman,}
\author[a]{D. Krapohl,}
\author[a]{H. Rahman,}
\author[b]{D. Chapman,}
\author[a,c]{R.H. Menk,}
\author[a]{and B. Norlin}

\affiliation[a]{Mid Sweden University,\\Sundsvall, Sweden}
\affiliation[b]{University of Saskatchewan,\\Saskatoon, Canada}
\affiliation[c]{Elettra Sincrotrone Trieste,\\Trieste, Italy}

\emailAdd{farangis.foroughi@miun.se}

\abstract{The transition toward renewable, fiber-based packaging requires an improved understanding of chemical modifications in high-yield pulps such as chemithermomechanical pulp (CTMP). Sulfonation uniformity is essential for the energy-efficient production of high-strength CTMP pulp. However, laboratory methods only measure total sulfur and cannot illustrate its distribution at the fiber level, which can be visualized using \textmu-XRF.
In this work, we present a laboratory \textmu-XRF system developed at Mid Sweden University and assess its capability to detect light elements in CTMP paper handsheets.
A \numproduct{32x32} point grid scan (\qtyrange{1.6}{1.6}{\mm^2} field of view, \qty{50}{\mu\m} step, 300 s/point) successfully resolved sulfur K\textalpha\ (\qty{2.31}{\keV}) and calcium K\textalpha\ (\qty{3.69}{\keV}) fluorescence without helium flushing. 
Comparative measurements at the Elettra synchrotron confirmed consistency of sulfur peak position and spatial distribution, with higher spectral resolution and signal-to-noise ratio. Histogram analysis using Wasserstein distance metrics demonstrated close agreement between datasets despite differing acquisition conditions. 
These results demonstrate that laboratory XRF can reproducibly detect and map sulfur in CTMP fibers under ambient conditions, providing a practical tool to complement synchrotron studies and supporting the development of energy-efficient, fiber-based packaging materials.}


\begin{document}
\maketitle
\flushbottom

\section{Introduction}
\label{sec:intro}

The transition from fossil-based plastics to renewable, fiber-based packaging materials has become increasingly important to reduce environmental impacts. High-yield pulps, particularly chemithermomechanical pulp (CTMP), are gaining attention as a sustainable raw material for paperboard and packaging applications. CTMP production relies on the impregnation of wood chips with sodium sulfite (\ce{Na_2SO_3}) to soften lignin and enable efficient fiber separation during refining \cite{rah2025}. The degree and uniformity of sulfonation critically influence pulp properties, including fiber bonding, bulk, and strength. However, sulfonation during impregnation is typically non-uniform, since outer chip layers absorb more sulfite (\ce{-SO_3^{-}}) than the core. Variations in wood chip size further contribute to uneven sulfonation and heterogeneous fiber properties. Achieving more even sulfonation is therefore a key research target for improving CTMP energy efficiency and final product quality \cite{ben1988}. 

To optimize such processes, it is necessary to measure the spatial distribution of sulfur at the fiber level. Advanced analytical tools, in particular, synchrotron \textmu-XRF, provide high spatial resolution and sensitivity, allowing clear detection of sulfur heterogeneity within single fibers. For example, recent studies at the Advanced Photon Source (Chicago, IL, USA) \cite{nor2023} and at Elettra (Trieste, Italy) \cite{rah2025} have mapped sulfur distributions in CTMP fibers, confirming strong variability in sulfonation and motivating further process optimization. While synchrotron measurements provide benchmark data, access is limited, and practical industry applications require laboratory-scale methods. 

Laboratory X-ray fluorescence (XRF) \cite{push2014} systems are attractive due to their potential to deliver non-destructive elemental maps under routine conditions. Previous work has demonstrated the use of laboratory XRF for paperboard quality control, especially in detecting calcium-based coatings \cite{nor2018}. However, detection of lighter elements such as sulfur under laboratory conditions has remained challenging, primarily due to low fluorescence yield, air absorption, and limited brilliance of conventional X-ray sources. Developments in optics and detector technology, including polycapillary focusing \cite{mac2010} and silicon drift detectors, open up possibilities for improved sensitivity at low energies.  

In this work, the capability of a laboratory-scale XRF system developed at Mid Sweden University to detect light elements in CTMP paper samples has been investigated. 
Handsheet samples (\qty{60}{\gram\per\m^2}) produced from CTMP pulp composed of \qty{30}{\percent} spruce and \qty{70}{\percent} birch were investigated, and elemental maps were obtained without helium flushing. 
Despite the limitations of laboratory sources, distinct sulfur K\textalpha\ (\qty{2.31}{\keV}) and calcium K\textalpha\ (\qty{3.69}{\keV}) peaks were successfully detected, enabling elemental mapping across the fiber network. To validate the laboratory results, comparative measurements were performed at the Elettra synchrotron facility. 

The study demonstrates that laboratory XRF, though limited in resolution and sensitivity compared to synchrotrons, can reliably detect and map sulfur in CTMP paper samples. This provides an accessible tool for fundamental studies of sulfonation heterogeneity and opens pathways for industrially relevant applications, complementing synchrotron measurements in advancing the development of energy-efficient, fiber-based packaging materials. 

\section{Materials and Methods}

\subsection{Sample preparation}
\begin{figure}[htbp]
\centering
\clipbox{0 0.4cm 0cm 0.5cm}{
    \includegraphics[width=.2\textwidth]{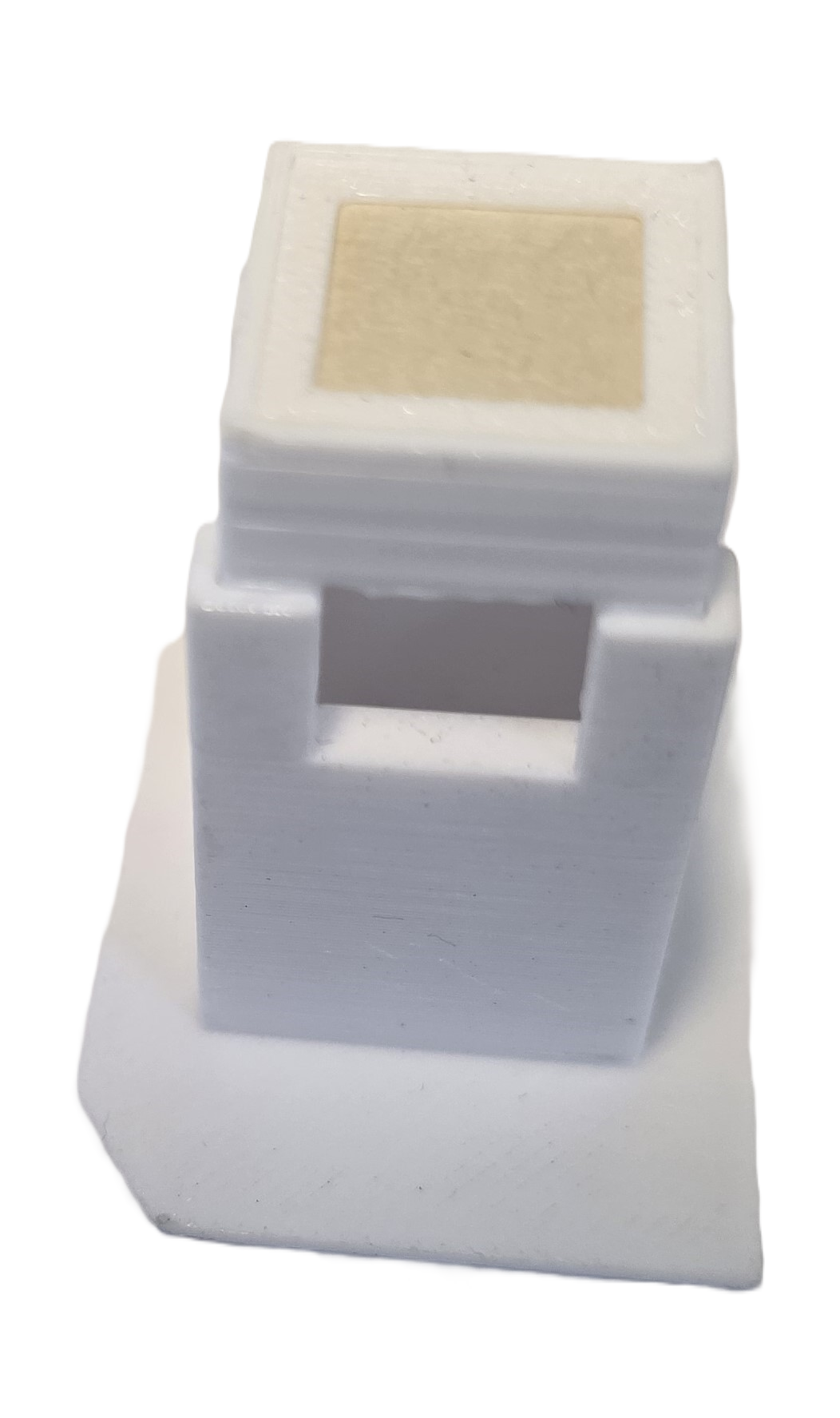}
}
\caption{Paper sample mounted on plastic holder for XRF measurement.\label{fig:sample_holder_lab}}
\end{figure}

Handsheets were prepared using the Rapid Köthen sheet former (Paper Testing Instruments; Pettenbach, Austria) following the guidelines outlined in ISO 5269-2:2004.
The handsheet samples were conditioned at room temperature prior to analysis, cut to \qtyproduct{10x10}{\mm^2}, and mounted on a custom 3D-printed plastic holder for XRF scanning (\cref{fig:sample_holder_lab}), ensuring stable positioning during the measurement.

Elemental mapping was performed across a \numproduct{32x32} point grid, corresponding to a field of view of \qtyproduct{1.6x1.6}{\mm^2}. The scan step size was \qty{50}{\um}, and the exposure time per point was  \qty{300}{\s}. This resulted in a total acquisition time of about 85 h, excluding additional stage overheads for positioning and data handling. The acquisition parameters were chosen to ensure sufficient counting for sulfur detection and to enable meaningful comparison with the synchrotron measurements. 

\subsection{Lab XRF Scanning System}

\begin{figure}[htbp]
\centering
\clipbox{1cm 0cm 0pt 0pt}{
    \includegraphics[width=0.47\textwidth, angle=180]{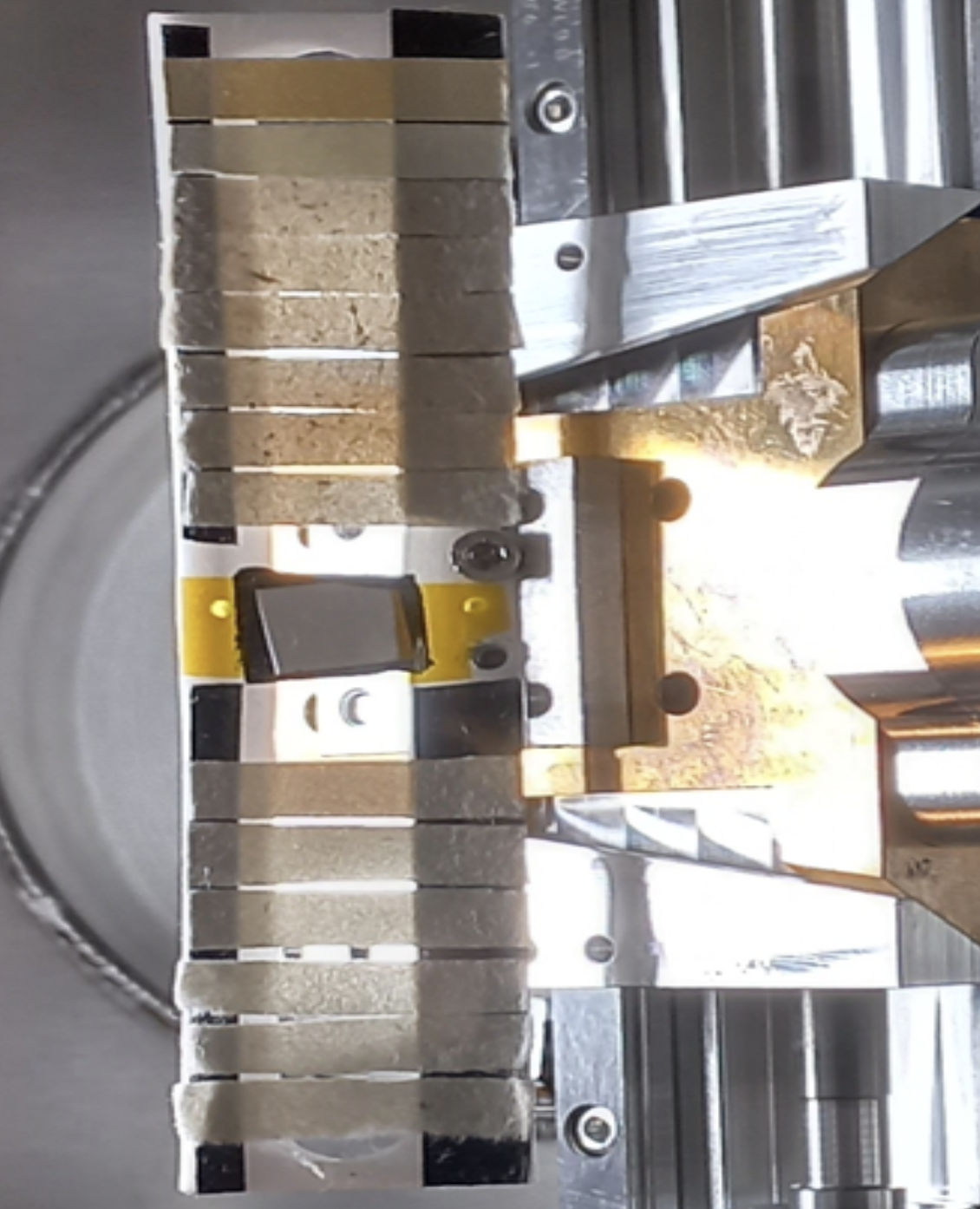}}
\clipbox{0pt 2cm 1cm 0pt}{%
  \rotatebox{-90}{\includegraphics[width=0.71\textwidth]{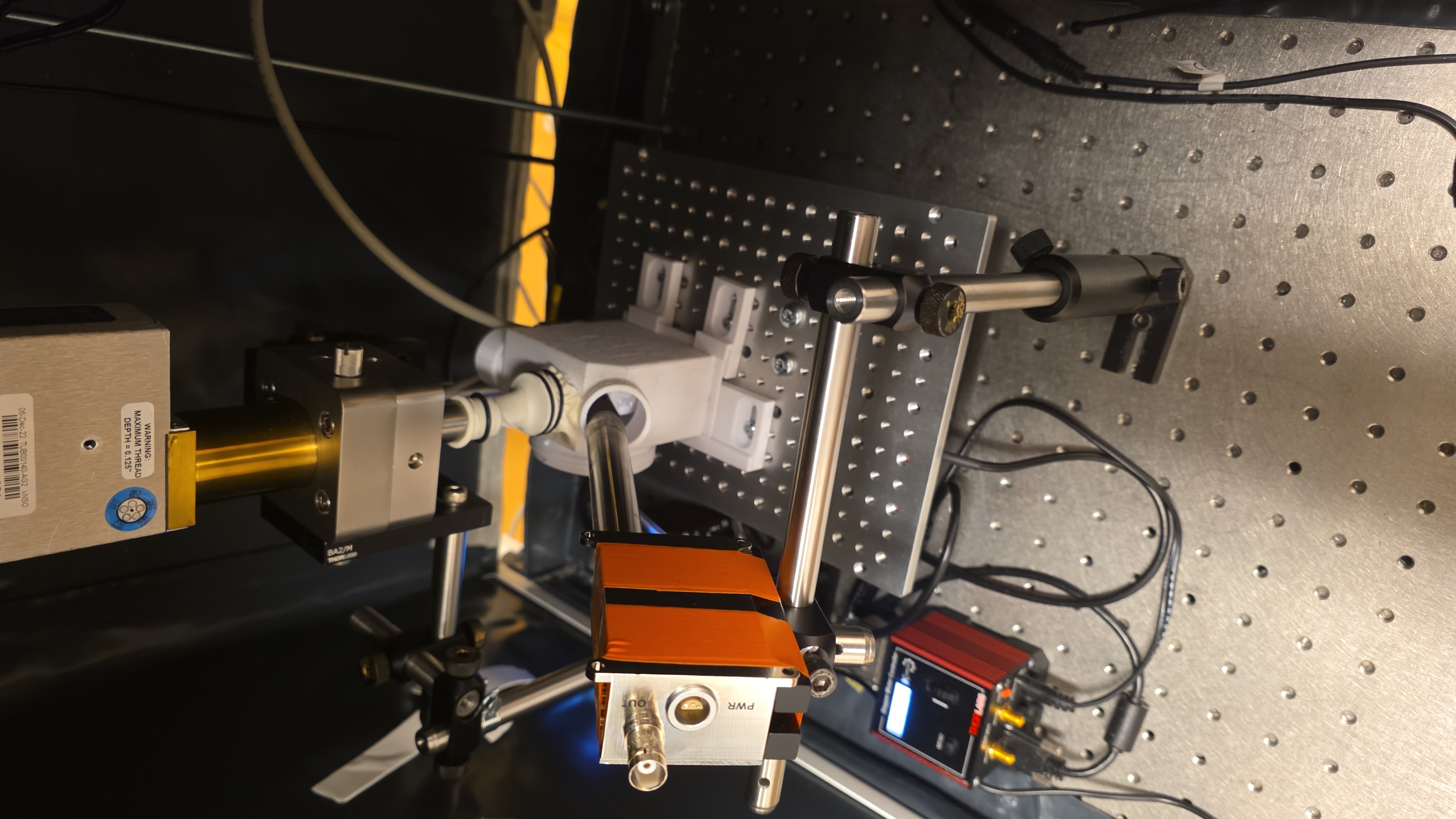}}%
}
\caption{XRF experiment setup. The left image shows the sample holder at the synchrotron. The right image shows the lab setup with the xy-stage controller, spectrometer, and source visible.  \label{fig:setup_lab}}
\end{figure}

The laboratory \textmu-XRF system consisted of a Moxtek Mini-Beam X-ray tube with a silver anode and polycapillary focusing optics, an Amptek XR-100 fast silicon drift detector (FastSDD) with a PX5 digital pulse processor. The sample holder was mounted inside a 3D-printed housing, intended for future experiments with helium flushing. The housing was fixed to an aluminium optics table, which was positioned on a motorised XY-stage (\cref{fig:setup_lab}, right).

The microfocus tube has a maximum operating power of \qty{50}{\kV}/\qty{12}{\W}. The integrated polycapillary optics provided a \qty{3.3}{\mm} focus and is capable of producing a nominal \qty{12}{\mm} focal spot. 
For the present measurements, the source was operated at \qty{12}{\kV} and \qty{900}{\micro\ampere},  providing sufficient excitation of the sulfur (\qty{2.31}{\keV}) and calcium (\qty{3.69}{\keV}) K\textalpha\ lines while keeping the incident energy low enough to reduce Compton scattering.

Fluorescence spectra were recorded using an Amptek XR-100 FastSDD with \qty{25}{\mm} active area, \qty{500}{\mm} thick silicon drift detector, and a beryllium (Be) entrance window.
The sample was positioned approximately \qty{3}{\mm} below the exit tip of the polycapillary lens, corresponding to the focal distance of the optic. 
The detector was positioned at an angle of \qty{\approx 45}{\degree} to the incident beam, with a \qty{\approx 45}{\degree} take-off angle relative to the sample surface.
The sample was mounted on a motorized XY-translation stage (Thorlabs LNR25ZFS), enabling controlled raster scanning. The entire setup was built on an optics table and placed in a lead-shielded box. 

\subsection{Synchrotron Measurements}
To benchmark the laboratory measurements, complementary \textmu-XRF experiments were performed at the Elettra synchrotron XRF beamline. 
The beamline provides radiation in the \qtyrange{2}{14}{\kilo\electronvolt} energy range. 
For the present study, the incident energy was tuned to \qty{4.3}{\keV} to optimize the excitation of the sulfur K\textalpha\ emission and reduce Compton background. 

The beam was focused to a spot size of approximately \qty{50}{\um} on the sample, which defined the effective spatial resolution of the measurement. Raster scanning was carried out with a step size of \qty{40}{\um}, smaller than the beam size, leading to oversampling. The beamline is equipped with ultra-high vacuum (UHV) chambers, a sample transfer system, and a seven-axis motorized stage for precise positioning. Fluorescence detection was performed with an XFlash 5030 SDD detector (Bruker Nano GmbH; Berlin, Germany) equipped with a zirconium collimator. The detector has a \qty{25}{\mm\squared} active area and an energy resolution of \qty{131}{\eV} (at Mn K\textalpha\ \qty{5.9}{\keV}). Measurements were conducted under vacuum conditions to minimize absorption losses of low-energy fluorescence. \Cref{fig:setup_lab} (left) shows the sample holder inside the vacuum chamber of the beamline. 

A side-by-side summary of the key experimental parameters for the lab and synchrotron setups is provided in \Cref{tab:i}.
\begin{table}[htbp]
\centering
\caption{Key experimental parameters of the laboratory \textmu-XRF system and the synchrotron XRF setup, highlighting differences in excitation, geometry, and sampling strategy.\label{tab:i}}
\smallskip
\begin{tabular}{>{\raggedright\arraybackslash}p{3cm}|>{\raggedright\arraybackslash}p{5.2cm}|>{\raggedright\arraybackslash}p{5.2cm}}
\hline
\textbf{Parameter} & \textbf{Laboratory setup (MIUN)} & \textbf{Synchrotron setup (Elettra)} \\
\hline
X-ray source & Moxtek Mini-Beam, Ag anode, XOS poly-capillary optic & Synchrotron radiation, crystal monochromator \\
Operating energy & \qty{12}{\kV}, \qty{900}{\micro\A} (broad spectrum) & \qty{4.3}{\keV} (monochromatic) \\  
Spot size  & \qtyrange{\sim 15}{20}{\mu\m} (polycapillary focus) & \qty{\sim 50}{\mu\m} (focused beam) \\
Step size & \qty{50}{\mu\m} & \qty{40}{\mu\m}\\
Environment & Air (no He purge) & Ultra-high vacuum (UHV) \\
Detector & Amptek XR-100 FastSDD, \qty{25}{\mm^2}, \qtyrange{125}{130}{\eV} at 5.9 keV & Bruker XFlash 5030 SDD, \qty{25}{\mm^2}, \qty{131}{\eV} at \qty{5.9}{\keV} \\
Detector geometry & \qty{45}{\degree} to beam, \qty{45}{\degree} to the sample plane & \qty{90}{\degree} to beam, \qty{45}{\degree} to the sample plane \\
Field of view (map) & \qtyproduct{1.6x1.6}{\mm^2} (\numproduct{32x32} grid) & \qtyproduct{1.6x1.6}{\mm^2} (\numproduct{32x32} grid) \\
Acquisition time & 300 s/pixel ($\sim$85 h, excl.\ overhead) & 5 s/pixel ($\sim$2.5 h) \\
\hline
\end{tabular}
\end{table}
\section{Results}
\subsection{Spectral Characterization}
\begin{figure}[htbp]
    \centering
    \includegraphics[width=0.8\textwidth]{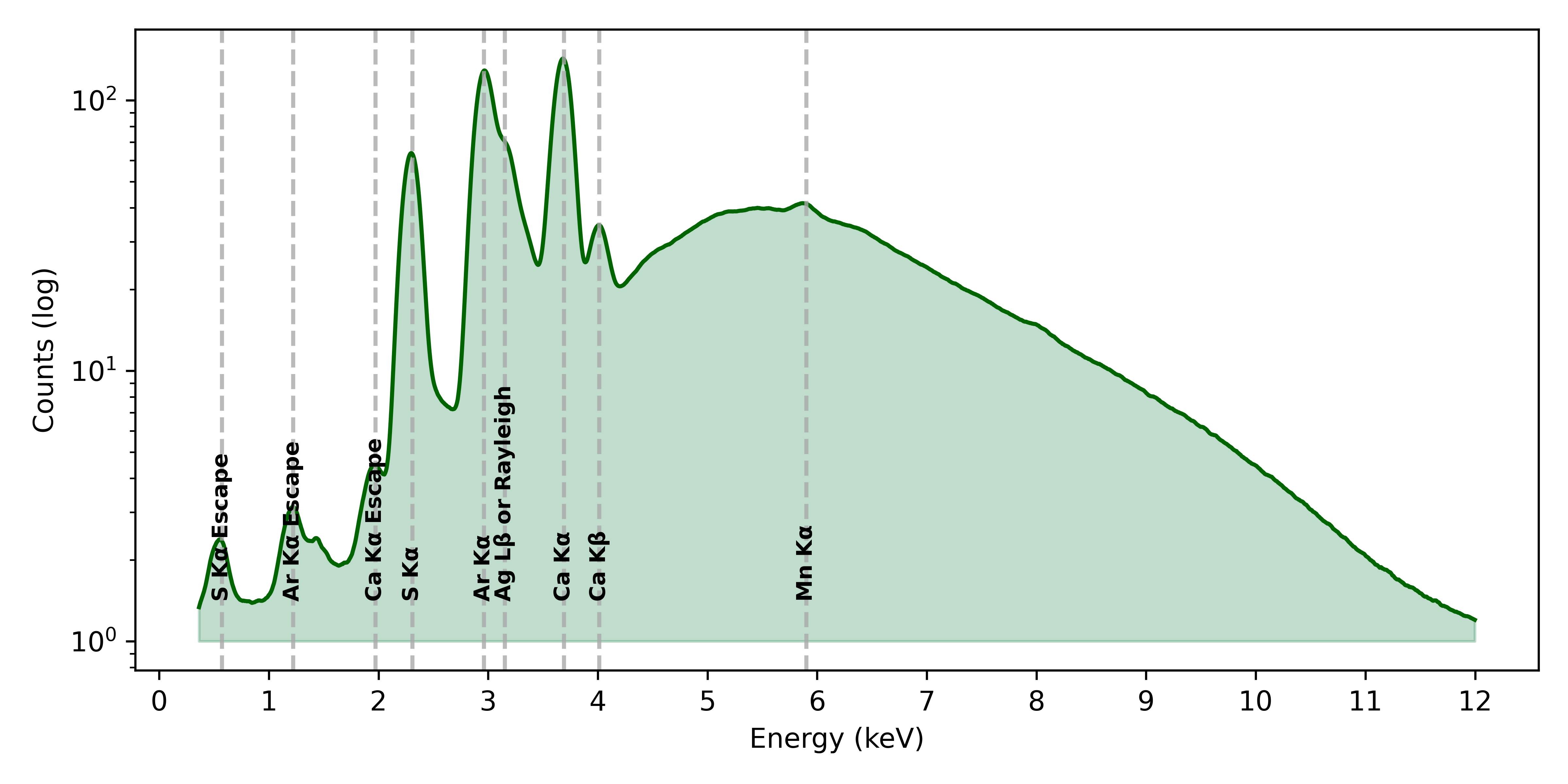}\\[0.5cm]
    \includegraphics[width=0.8\textwidth]{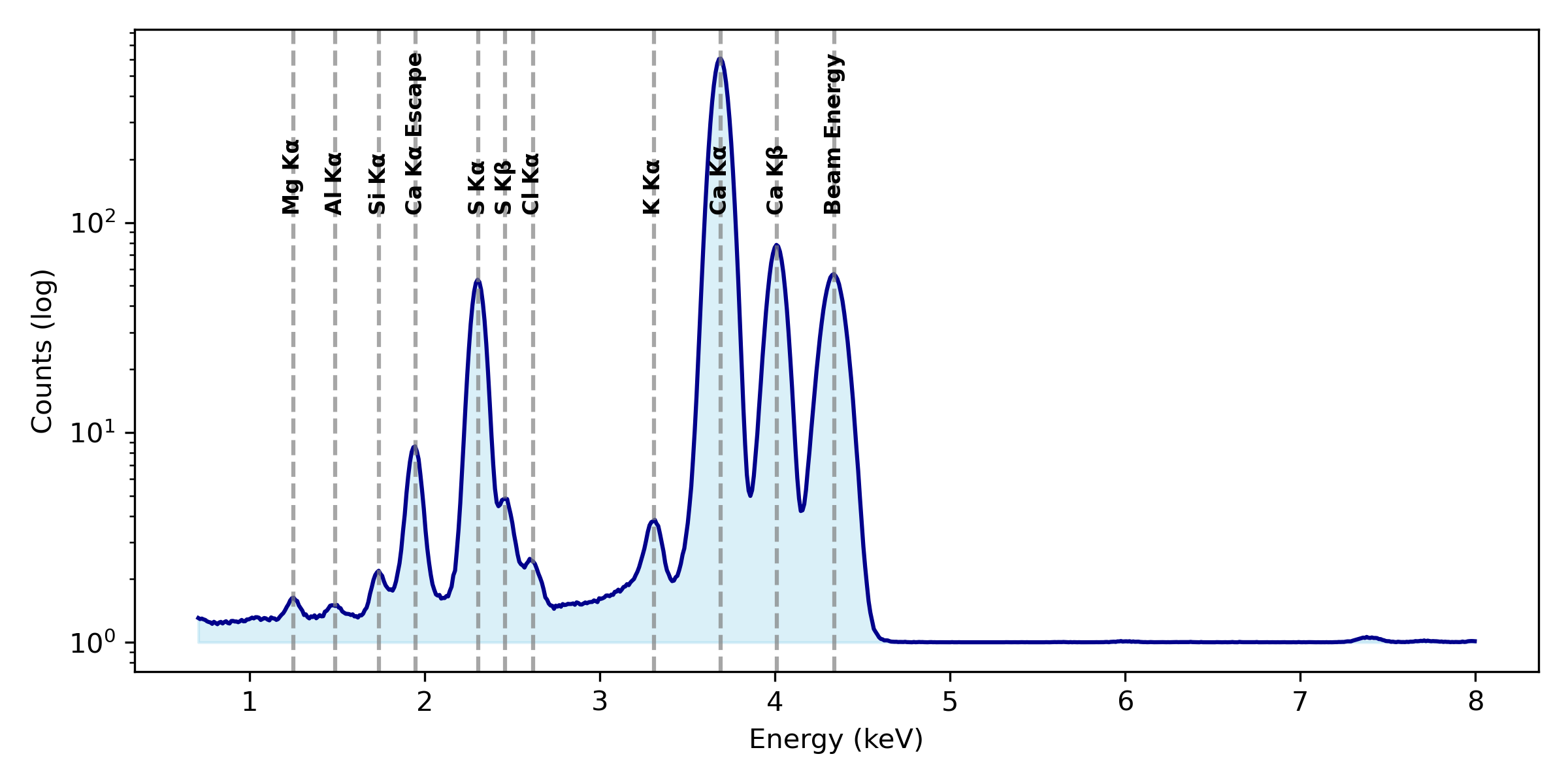}
    \caption{Comparison of XRF spectra. Top: full spectra from the lab.
             Bottom: full spectra from the synchrotron.}
    \label{fig:spectra_comparison}
\end{figure}

\Cref{fig:spectra_comparison} shows the average XRF spectra obtained from the lab and synchrotron measurements. Both spectra reveal multiple characteristic emission lines, including sulfur K\textalpha\ (\qty{\approx 2.30}{\keV}), calcium K\textalpha\ (\qty{\approx 3.69}{\keV}), as well as weaker features from other elements, scattering, and escape peaks. The synchrotron spectrum exhibits sharper and more distinct elemental peaks, reflecting the superior energy resolution and stability of the synchrotron setup. In contrast, the lab spectrum shows a broader background, consistent with the polychromatic source. Despite these differences, the elemental fingerprints are consistent across the two datasets, confirming that the same chemical species were probed. 

\begin{figure}[htbp]
\centering
\includegraphics[width=.45\textwidth]{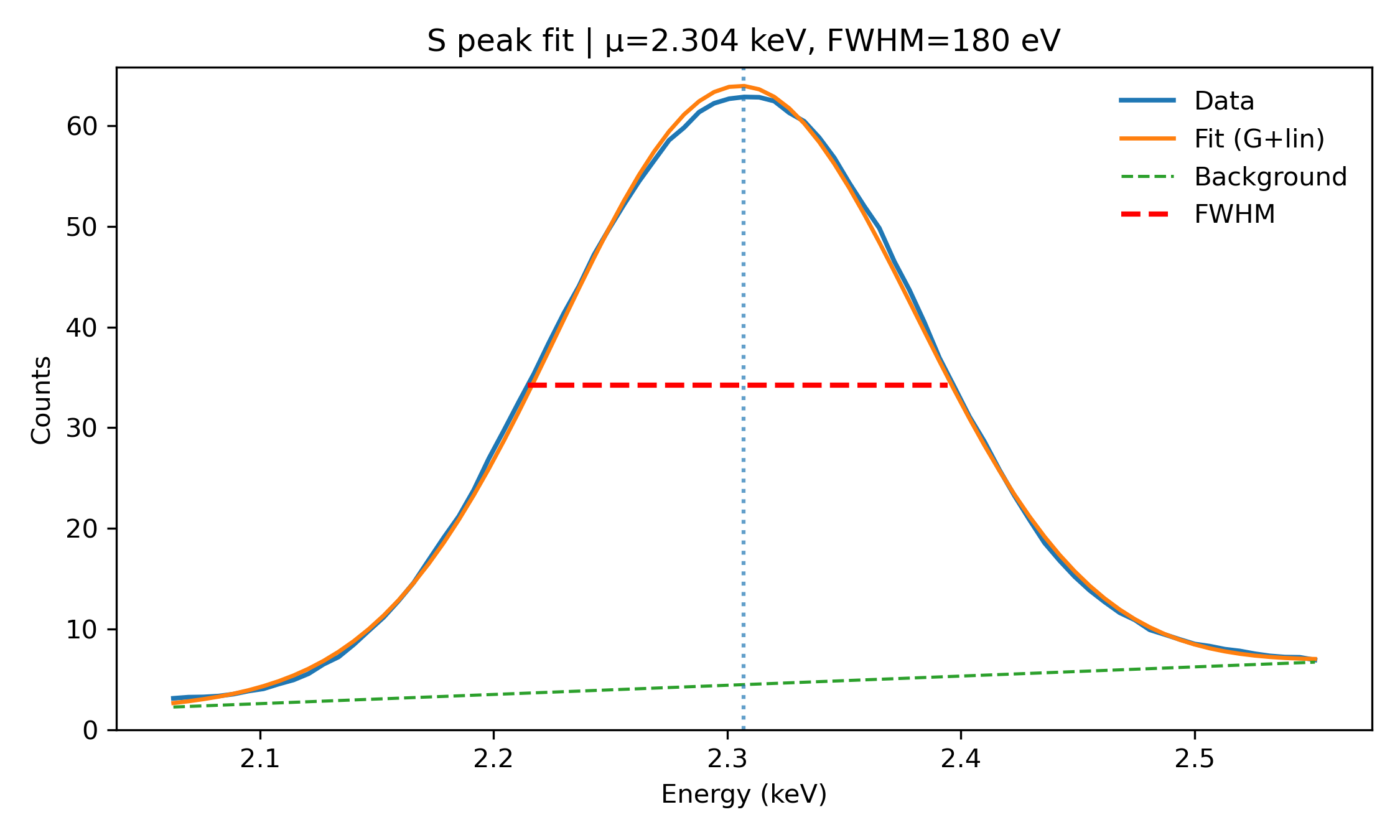}
\includegraphics[width=.49\textwidth]{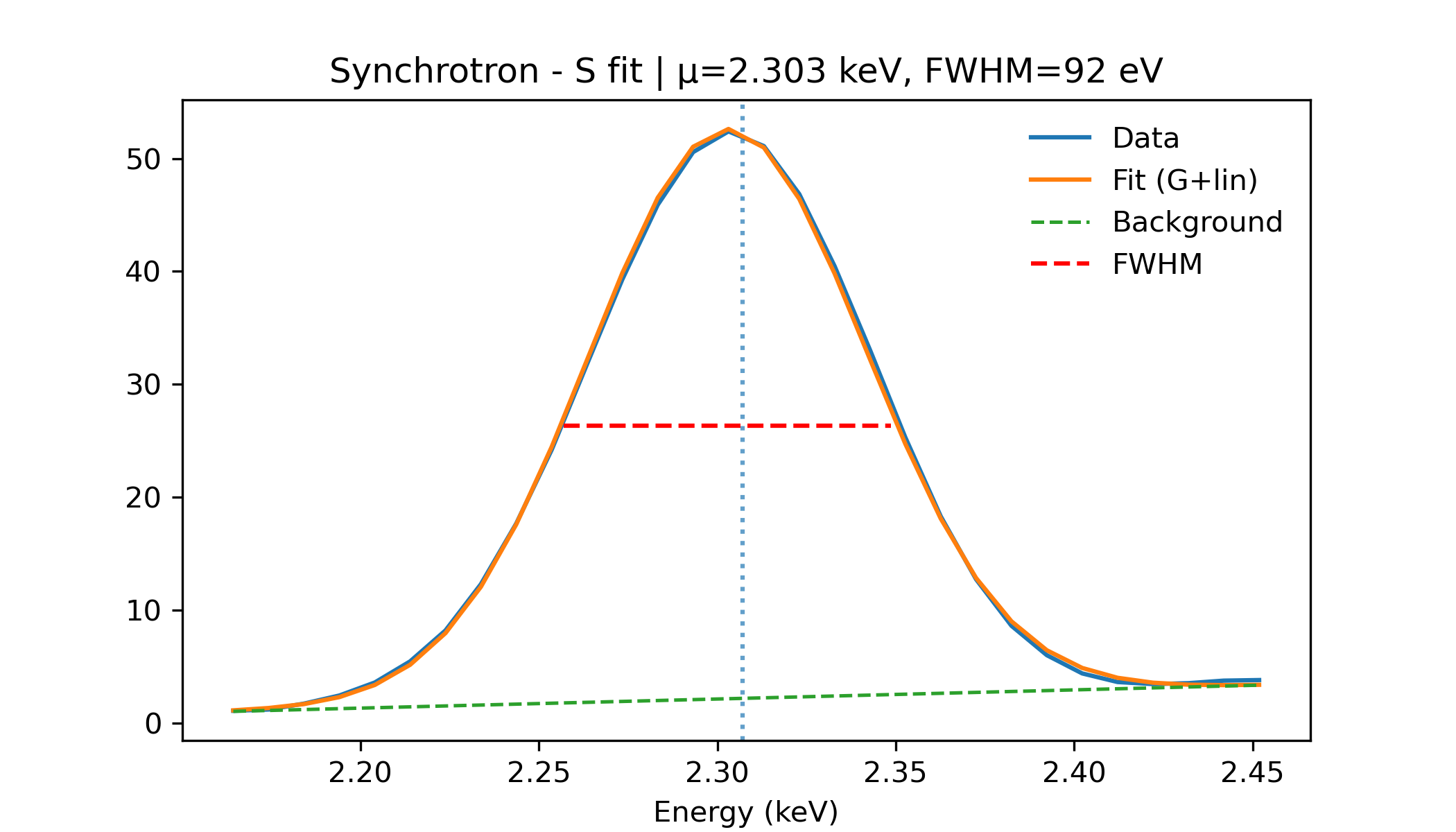}
\caption{Sulfur peak fits from the laboratory (left) and synchrotron (right) datasets. 
Both spectra show the expected emission at \qty{\approx 2.30}{\keV}, fitted with a Gaussian plus linear background model.\label{fig:sulfur_peak_fit}}
\end{figure}

\Cref{fig:sulfur_peak_fit} presents detailed fits of the sulfur K\textalpha\ emission lines for both datasets. 
A Gaussian peak with a linear background was fitted to extract peak positions. The sulfur K\textalpha\ line was consistently observed at \qty{\approx 2.30}{\keV} in both datasets, confirming reproducibility across measurement platforms. 
The synchrotron spectrum exhibited a narrower profile with a fitted FWHM of \qty{92}{\eV}, corresponding to a higher energy resolution. 
In contrast, the laboratory spectrum showed a broader peak with a FWHM of \qty{180}{\eV}, reflecting its lower energy resolution. 
As the spectra were acquired from different regions of the paper sample, the intensity differences may reflect genuine local variation in sulfur content across the heterogeneous fiber network, though effects from source properties and detector geometry cannot be excluded. 
Nevertheless, the agreement in peak energy between the two datasets confirms that the laboratory setup reliably captures the sulfur signal in line with the synchrotron reference. 
The ability to detect sulfur in air further demonstrates the sensitivity of the setup to low-energy fluorescence lines, which is particularly important for CTMP research, where mapping sulfur distribution at the fiber scale is crucial for achieving uniform sulfonation.

\subsection{Comparison of Laboratory (Lab) and Synchrotron Sulfur Maps}
To assess the consistency between lab-based and synchrotron-based XRF imaging, we compared sulfur (K\textalpha) distributions obtained from two datasets. Although the lab and synchrotron scans were acquired from different regions of the paper sample, both datasets were processed using an identical normalization procedure (row-wise normalization of pixel intensities). It should be noted that due to the heterogeneous nature of sulfonation and fiber distribution, structural variations between regions are expected. The comparison, therefore, provides an assessment of overall reproducibility rather than a pixel-wise validation. 

\subsubsection{Spatial Sulfur Maps}
\begin{figure}[htbp]
\centering
\includegraphics[width=.9\textwidth]{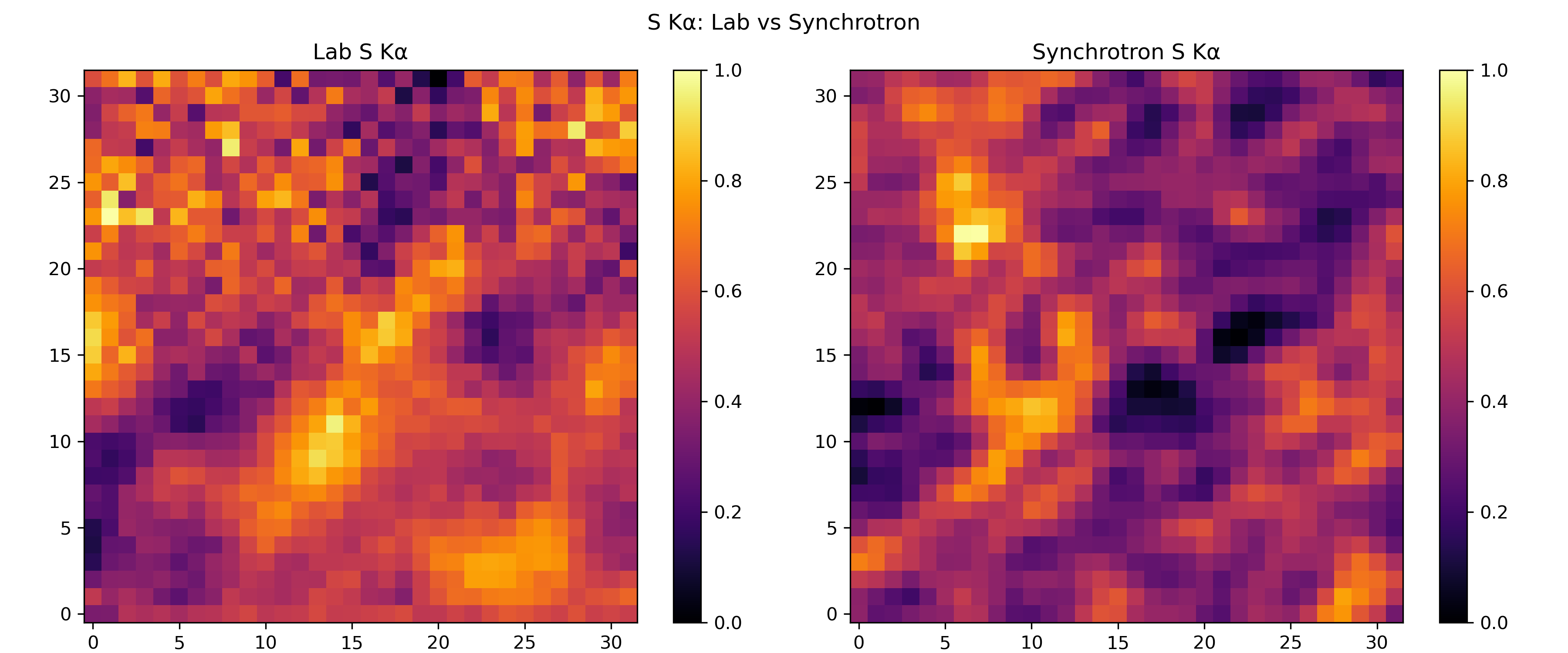}
\caption{Sulfur (S K\textalpha ) maps for the lab and synchrotron datasets. The color bar indicates relative sulfur intensity.\label{fig:sulfur_map}}
\end{figure}

\Cref{fig:sulfur_map} presents the spatial distribution of sulfur (SK\textalpha) obtained from both the lab (left) and synchrotron (right) datasets. 
Both maps reveal a heterogeneous distribution of sulfur, with localized high-intensity domains that can be associated with fiber-rich regions. Although the detailed spatial patterns differ between the two measurements (as expected given the regions of the samples are not identical) overall presence of sulfur-rich structures is consistent across both datasets, supporting the reproducibility of the underlying material features. 

It should be noted that the lab acquisition was affected by detector overheating, which introduced systematic row-wise variations in intensity (Section A, \cref{fig: detector overheating}, left). To mitigate this artifact, a row-normalization procedure was applied to both datasets, ensuring comparable treatment of the images. 
The synchrotron dataset was further normalized to the incident photon flux, measured by a monitoring diode, in order to correct for beam current variations during the scan.
Finally, to facilitate direct comparison of relative signal strength, the maps were globally normalized to the [0,1] scale. 
The color bar in \cref{fig:sulfur_map} therefore represents relative sulfur intensity, where 0 corresponds to the minimum and 1 to the maximum value within each dataset. 

\subsubsection{Intensity Distribution}
\begin{figure}[htbp]
\centering
\includegraphics[width=.4\textwidth]{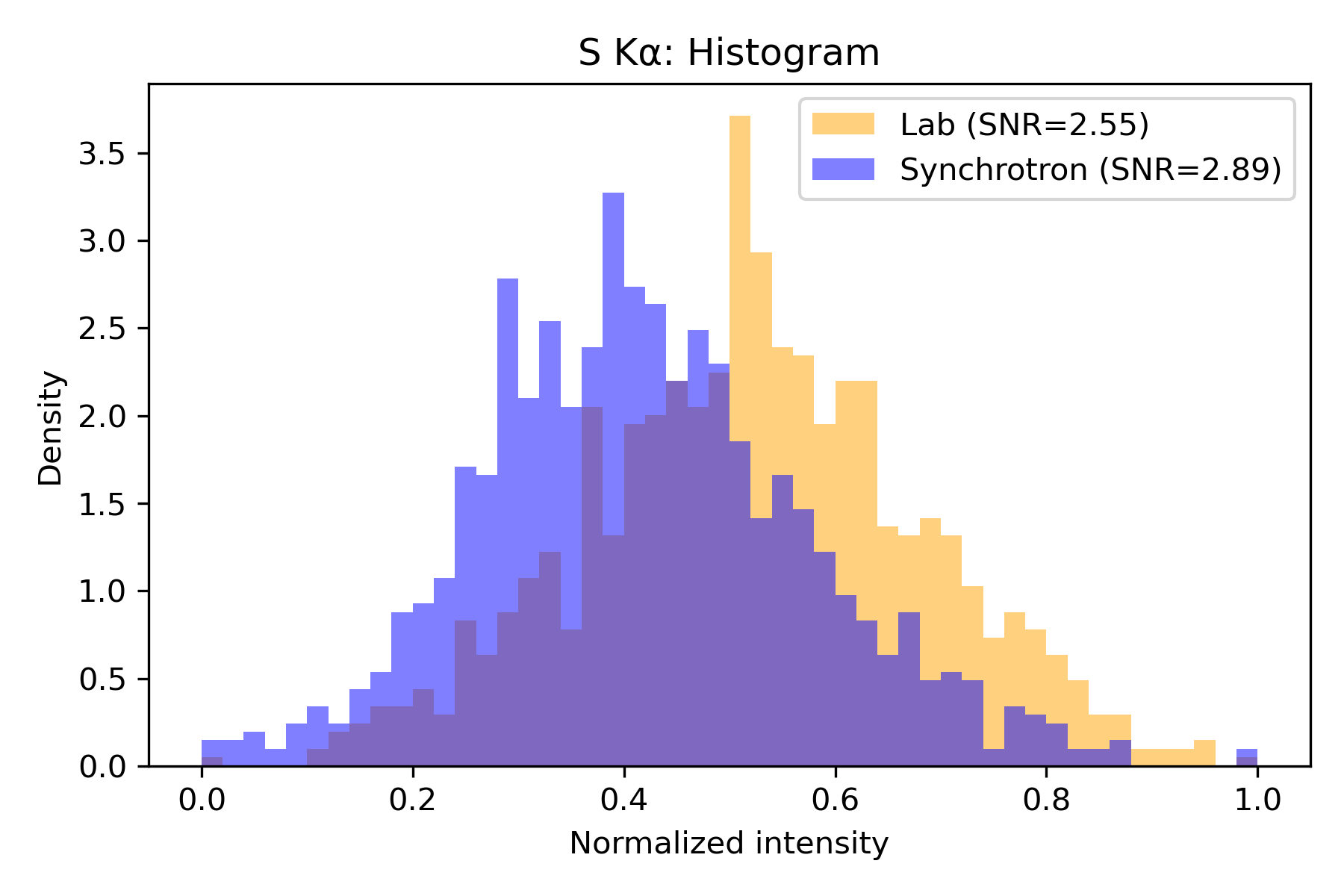}
\includegraphics[width=.4\textwidth]{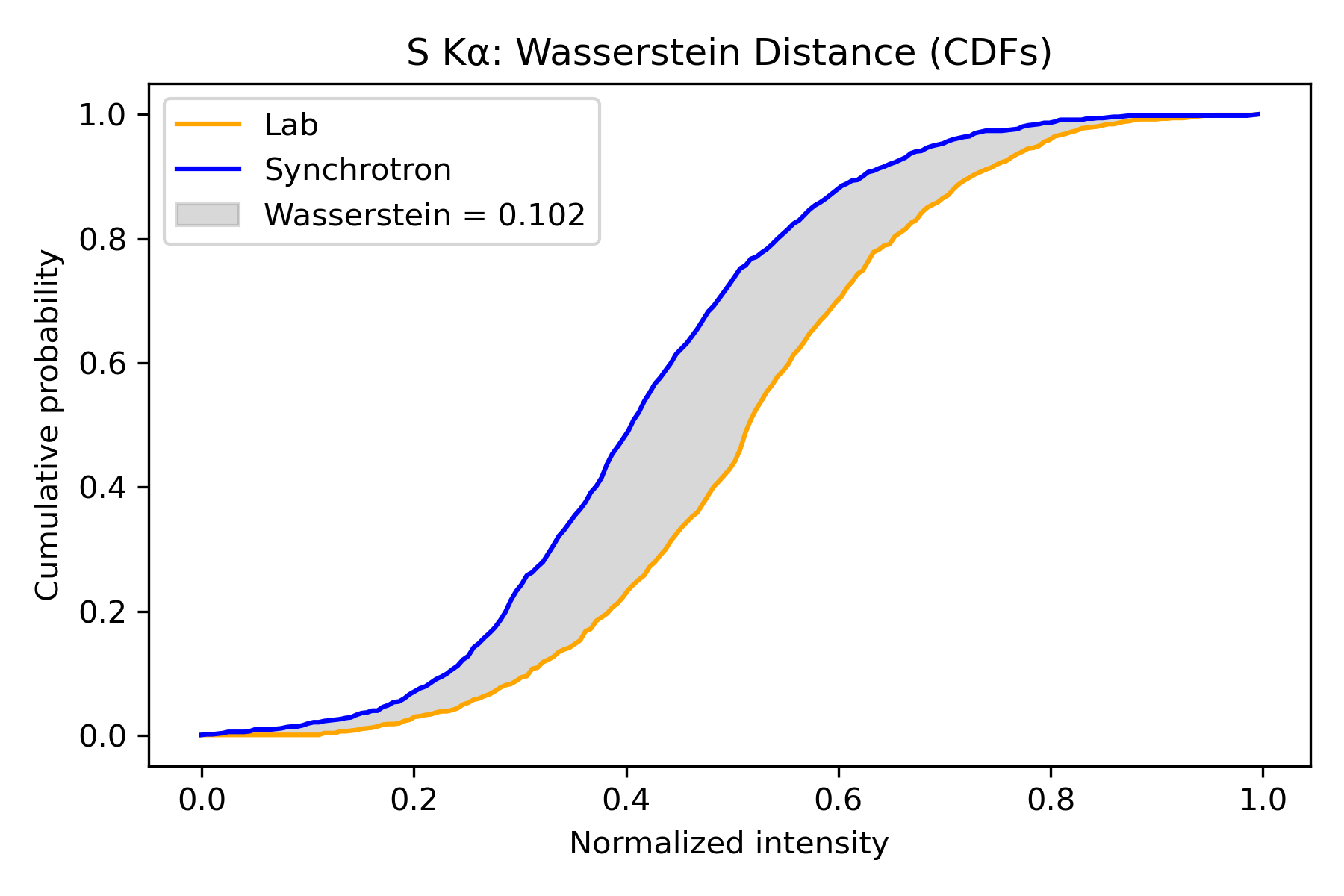}
\caption{Sulfur $(S K \alpha)$ Normalized intensity distribution for the lab and synchrotron datasets (left). Cumulative distribution functions (CDFs) of the normalized sulfur intensities with shaded region indicating their Wasserstein distance (right).\label{fig:sulfur_pix_histogram}}
\end{figure}

\Cref{fig:sulfur_pix_histogram} shows the normalised intensity histograms of the lab and synchrotron sulfur maps. 
While both datasets display broad, overlapping distributions characteristic of heterogeneous sulfur content, systematic differences are apparent. The lab data are centered at a slightly higher mean intensity with a narrower spread, indicating a more uniform distribution of pixel values. In contrast, the synchrotron data extend further toward lower intensities and exhibit a broader variance, reflecting greater local intensity variations. 
The signal-to-noise (SNR) is higher for the synchrotron map (2.89) compared with the lab map (2.55), consistent with higher photon flux and improved photon counting statistics of the synchrotron source. 
This higher SNR accounts for the ability of the synchrotron data to resolve finer-scale intensity variations that remain obscured in the lab data. 
The similarity of the two datasets was further assessed using the Wasserstein distance between their cumulative distribution functions. 
Unlike simpler metrics such as variance or mean comparison, the Wasserstein distance considers the full shape of both distributions, measuring the minimal "cost" of transforming one into the other. 
This makes it particularly well-suited for comparing empirical histograms of photon counts, where differences may arise from both shifts in mean intensity and changes in spread~\cite{gib2002}. 
The relatively small value (0.102) confirms that, despite differences in spread and spatial features, both measurements capture closely related sulfur intensity distributions. This indicates that both measurements are consistent in their overall depiction of sulfur content, with the synchrotron providing enhanced contrast due to its superior photon statistics. 

\section{Conclusion}
This study demonstrates that laboratory \textmu-XRF can be used to detect and map sulfur in CTMP paper handsheets under ambient conditions, without the need for vacuum or helium flushing. Despite the inherently lower brilliance of laboratory sources compared to synchrotrons, the system successfully resolved sulfur fluorescence peaks and produced elemental maps consistent with synchrotron measurements. The comparison highlighted differences in spectral resolution, which is expected due to source characteristics and photon flux, but confirmed reproducibility in peak energy and overall distribution of sulfur. The histogram and Wasserstein distance analyses further showed that both datasets captured closely related intensity distributions, highlighting the robustness of the laboratory system.

The total acquisition time of \qty{\sim 85}{\hour} per map remains a practical limitation for routine use, largely due to the long exposure times required for sulfur detection.
Nevertheless, ongoing advances in source brilliance, detector efficiency, and acquisition strategies are expected to reduce scan times significantly. 
Overall, the results establish laboratory XRF as an accessible method that translates synchrotron capabilities into the laboratory setting for fundamental studies of light element composition in fiber-based materials. 
As demonstrated with sulfonation heterogeneity in CTMP fibers, this capability enables industrially relevant applications in pulp and paper research.

\appendix

\renewcommand{\thefigure}{A\arabic{figure}}
\setcounter{figure}{0} 

\section{Recovery of Sulfur Maps from Detector Overheating Artifacts}
During the lab measurements, the detector was affected by overheating, which introduced systematic artifacts in the sulfur maps. 
\begin{figure}[htbp]
\centering
\includegraphics[width=.4\textwidth]{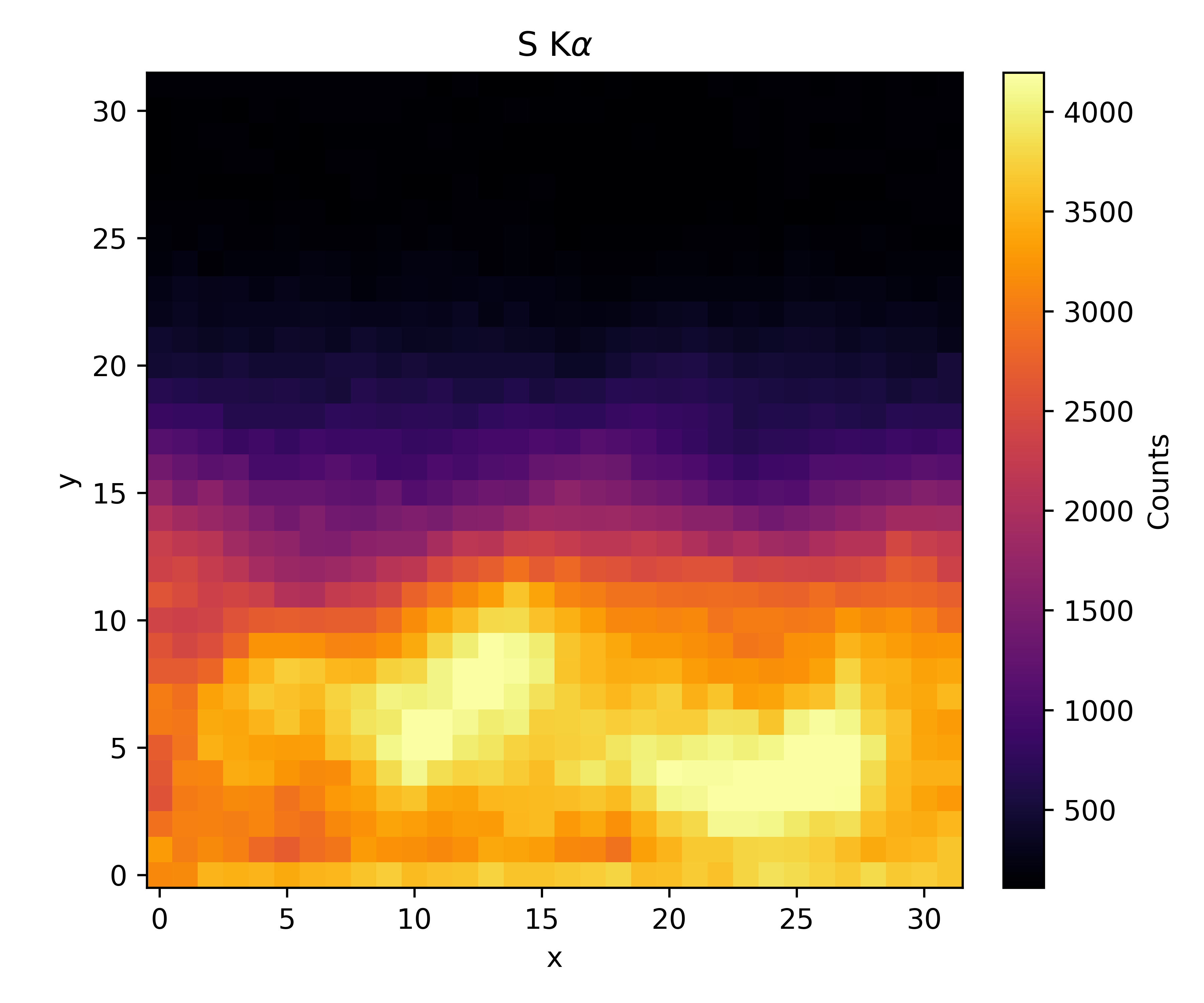}
\includegraphics[width=.4\textwidth]{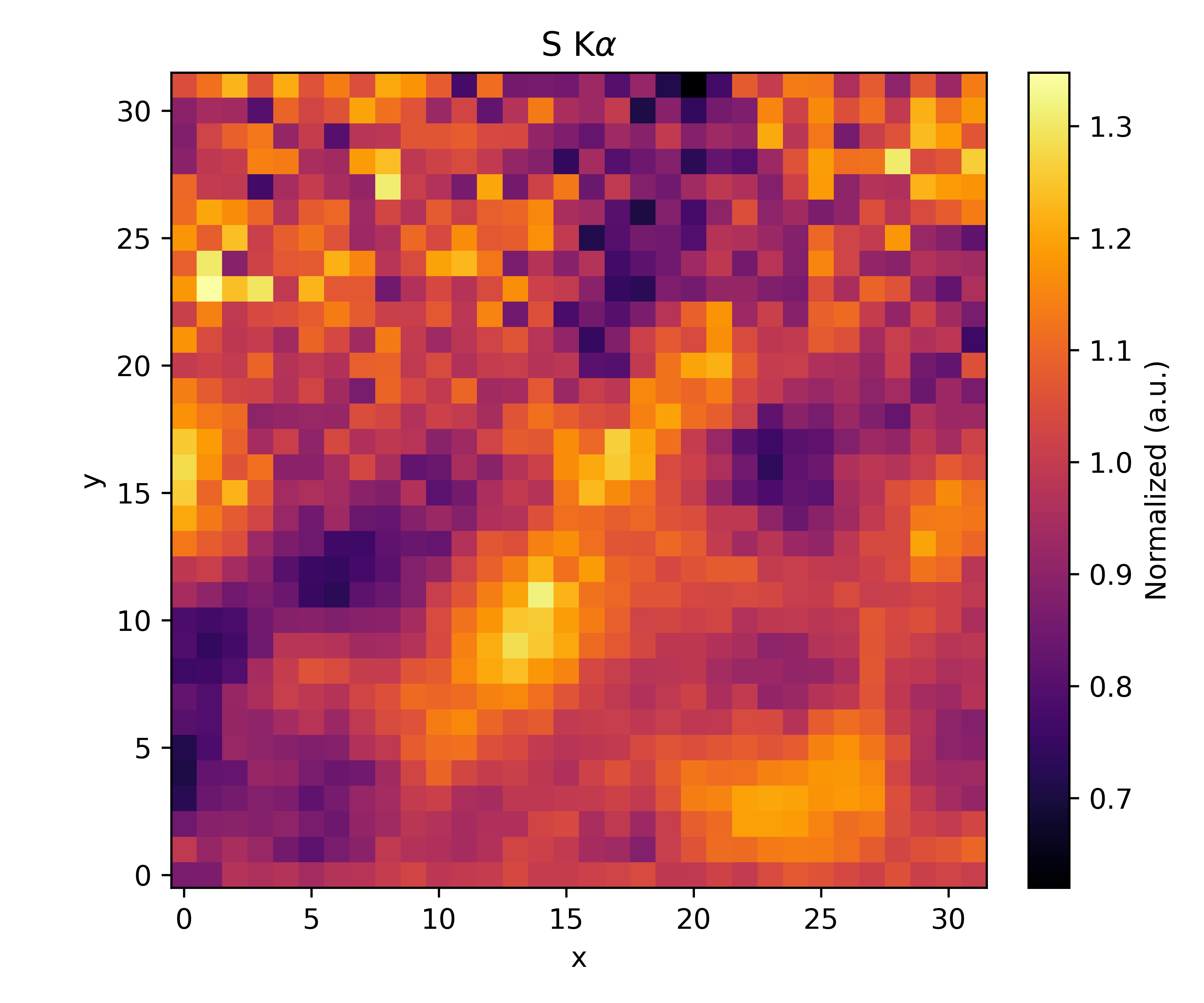}
\caption{Sulfur $(S K \alpha)$ map showing signal suppression in the upper half due to detector overheating (left). Row-normalized Sulfur $(S K \alpha)$ map restores sulfur heterogeneity across the field of view. (right).\label{fig: detector overheating}}
\end{figure}
To investigate further, spectra were examined from the apparently corrupted upper half (\cref{fig: detector overheating}, left). The spectra contained clear elemental peaks, including sulfur, indicating that the signal was not lost but instead suppressed by a gradual artifact. Based on this observation, a row-wise normalization procedure was implemented to correct the data, as illustrated in the \cref{fig: detector overheating} (right).

\acknowledgments
We acknowledge the support of the Elettra Synchrotron, co-funded by the NEPHEWS project (Grant Agreement No. 101131414, Horizon Europe). We also thank beam scientist Giuliana Aquilanti for her invaluable technical assistance. We further acknowledge the assistance and support with equipment provided by the MAX IV Laboratory.



\bibliographystyle{JHEP}
\bibliography{IWORID2025-FarangisForoughi}
\end{document}